\documentclass[twocolumn,prb,superscriptaddress]{revtex4-2} 
\usepackage{bm}
\usepackage{mathtools}
\usepackage{extarrows}
\usepackage{enumitem}
\usepackage{graphicx}
\usepackage{color,xcolor}
\usepackage{amsmath}
\usepackage{amssymb}

\newcommand{\Sec}[1]{Sec.\,\ref{#1}}
\newcommand{\App}[1]{Appendix\,\ref{#1}}

\newcommand{\gler}{\mbox{\tiny $\gtrless$}}

\newcommand{\T}{\mbox{\tiny Tot}}

\newcommand{\chain}{\rm cha}
\newcommand{\probe}{\rm pro}
\newcommand{\inter}{\rm int}

\newcommand{\greater}{\mbox{\tiny$>$}}
\newcommand{\lesser}{\mbox{\tiny$<$}}
\newcommand{\w}{\omega}
\newcommand{\ti}{\tilde}
\newcommand{\nl}{\nonumber \\}

\newcommand{\wti}{\widetilde}

\newcommand{\be}{\begin{equation}}
\newcommand{\ee}{\end{equation}}
\newcommand{\bsube}{\begin{subequations}}
\newcommand{\esube}{\end{subequations}}
\newcommand{\Eq}[1]{Eq.\,(\ref{#1})}
\newcommand{\Eqs}[1]{Eqs.\,(\ref{#1})}
\newcommand{\Fig}[1]{Fig.\,\ref{#1}}

\newcommand{\dg}{\dagger}
\newcommand{\la}{\langle}
\newcommand{\ra}{\rangle}

\renewcommand{\d}{{\rm d}}

\newcommand{\tL}{0}
\newcommand{\tR}{1}

\begin{document}

\title{
Local thermal probe in a one-dimensional chain: 
An efficient dissipaton-based approach
}

\author{Hao-Yang Qi}
\affiliation{State Key Laboratory of Precision and Intelligent Chemistry, University of Science and Technology of China, Hefei, Anhui 230026, China}

\author{Zi-Fan Zhu}
\affiliation{Hefei National Laboratory,
 University of Science and Technology of China, Hefei, Anhui 230088, China}
 \affiliation{Hefei National Research Center for Physical Sciences at the Microscale and Department of Chemical Physics, University of Science and Technology of China, Hefei, Anhui 230026, China}

\author{Yao Wang}
 \email{wy2010@ustc.edu.cn}
\affiliation{Hefei National Laboratory,
University of Science and Technology of China, Hefei, Anhui 230088, China}
\affiliation{Hefei National Research Center for Physical Sciences at the Microscale and Department of Chemical Physics, University of Science and Technology of China, Hefei, Anhui 230026, China}

\author{Rui-Xue Xu}
\email{rxxu@ustc.edu.cn}
\affiliation{State Key Laboratory of Precision and Intelligent Chemistry, University of Science and Technology of China, Hefei, Anhui 230026, China}
 \affiliation{Hefei National Laboratory,
 University of Science and Technology of China, Hefei, Anhui 230088, China}
 \affiliation{Hefei National Research Center for Physical Sciences at the Microscale and Department of Chemical Physics, University of Science and Technology of China, Hefei, Anhui 230026, China}

 \author{YiJing Yan}
 \affiliation{Hefei National Research Center for Physical Sciences at the Microscale and Department of Chemical Physics, University of Science and Technology of China, Hefei, Anhui 230026, China}

\date{\today}

\begin{abstract}
We study a system consisting of an infinite one-dimensional molecular chain and a locally coupled probe. Starting from the Hamiltonian of the chain–probe composite and the corresponding spectral densities, we evaluate the heat current between the probe and the chain. For this purpose, we develop a dissipaton-based quantum approach that is fully nonperturbative and non-Markovian. The dissipaton algebra yields a set of hierarchically coupled equations of motion for the dissipaton moments, with cross-tier connections in an iterative manner if higher-order chain–probe interactions are included. Numerical results demonstrate the effects of temperature, frequency, onsite energy modification and higher-order couplings on heat transport. This work provides a general framework for thermal transport and other properties in locally probed systems and can be straightforwardly extended to higher-dimensional materials and electronic transport problems with strong many-body effects.
\end{abstract}

\maketitle

\section{Introduction}

In recent years it is of increasing interest in the thermal transport property through a molecular chain.\cite{Dub11131,Li121045} In the exploration, people may exert an impurity or a probe on a particular site of the chain where in reality, the trapping interaction may involve certain many-body effect or anharmonicity.\cite{Ahm061159,Che10134103,Ber10063415,Alo13023001,Zha16125503} 
Similar studies also include the 1D ultracold Bose gas,\cite{Sch26290} 
1D ballistic phonon waveguide,\cite{Tav184287} and model proteins.\cite{Ati01505,Ric09464,Wan23214105}
The influence and mechanism caused by the many-body  interaction or anharmonicity, together with the quantum effect, on the thermal property
and the subsequent function of related materials are thus under investigation.\cite{Dub11131,Li121045} Actually it has been found that anharmonicity can play a crucial role in the regulation and control of material functions.\cite{Reg98232,Che08105501,Li151063}

On the other hand, accurate and efficient theoretical study on the nanoscale quantum thermal properties with anharmonicity remains challenging. 
In past years, people have adopted
classical Langevin or Nosé--Hoover thermostat approaches,\cite{Gen002381,Li05104302,Lan13052126}
nonequilibrium molecular dynamics,\cite{Saa14134312}
nonequilibrium Green's function or self-consistent approaches,\cite{Min06125402,Lui12245407,He16155411,He18195437,Lee18205447,Guo20195412,Guo21174306}
and a recent mode coupling theory combined with the Monte Carlo simulation.\cite{Cas25094306}
The theoretical method development is also in fundamental analogy to particle dynamics with many-body interactions in crystals, interfaces and mesoscopic devices.

\begin{figure}[t]
\includegraphics[width=\columnwidth]{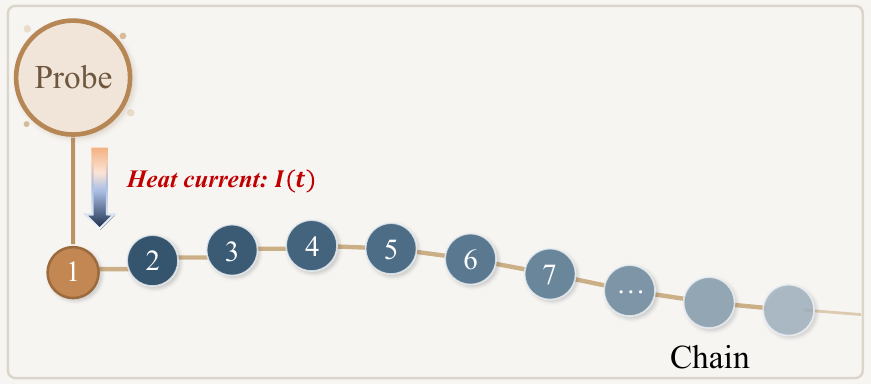}
\caption{Schematic illustration of the model: probing a one-dimensional vibrational chain  via the heat current caused by a local chain–probe interaction and onsite energy modification at a single site.
}\label{fig1}
\end{figure}

In this work, we apply the dissipaton-equation-of-motion (DEOM) approach\cite{Yan14054105,Zha15024112,Xu151816,Wan20041102,Wan22170901} to study the local thermal property of a 1D chain with a probe (cf.\,\Fig{fig1}). 
DEOM adopts {\em dissipatons} as quasi-particles for the characteristic statistical properties, and was originally established as an exact method for open quantum systems in Gaussian environments.\cite{Yan14054105,Zha15024112,Xu151816}  
 DEOM has been shown to be convenient and valid to include nonlinear, higher-order system-environment couplings for such as many-body interaction or anharmonicity through the dissipaton algebra.\cite{Xu18114103,Su23024113,Su254107,Zhu25234103,Zhu2667}  In this work, we will further develop the DEOM formalism for dissipaton moments yielding the form composed of normal c-numbers
as variables, instead of density matrices, in the coupled differential equations, without any further
approximations/restrictions. Therefore the method is highly efficient, meanwhile maintaining all the quantum and non-Markovian, non-perturbative effects. In the final hierarchically constructed DEOM, anharmonicity is dealt with through cross-tier connections in a recursive manner to account for higher-order interactions.

The paper is organized as follows. In \Sec{sectheo}, we present the model Hamiltonian and the corresponding spectral densities (\Sec{theoa}), with more details given in \App{appa}, as well as the form of the DEOM established for dissipaton moments 
(\Sec{theob}). In \App{appb},
the DEOM formalism for moments is detailed in the condition of bi-linear chain-probe coupling in \App{appB1}, which is shown to be consistent with the analytic results from fundamental quantum mechanics in \App{appB2}. 
Numerical demonstration and discussion are presented in \Sec{secnum}. Finally we summarize the paper in \Sec{secsum}. Throughout the paper, we set the Planck constant and Boltzmann constant as $\hbar=1$ and $k_B=1$, and $\beta_{\alpha}=1/T_{\alpha}$, with $\alpha=0$ and $1$ representing probe and chain, respectively.

\section{Theory}\label{sectheo}
\subsection{Hamiltonian and  spectral density}
\label{theoa}

The composite consists of a chain and a probe (cf.\,\Fig{fig1}), whose 
Hamiltonian reads
\be 
H_{\T}=H_{\chain}+H_{\probe}+H_{\inter}+H'.
\ee
The chain is composed of $N$ oscillators,  whose Hamiltonian reads
\begin{align}\label{Hchain}
H_{\chain}=\sum_{u=1}^{N}\frac{\hat{p}^{2}_{u}}{2m}+\frac{\kappa}{2}\sum_{u=1}^{N-1}(\hat{q}_{u+1}-\hat{q}_{u})^2.
\end{align}
The probe is described as a phonon bath 
\be 
H_{\probe}=\sum_{j}\epsilon_{j} \hat b_{j}^{\dg}\hat b_j.
\ee
We consider a long chain effectively with the limit $N\rightarrow \infty$ and apply the periodic boundary condition in dealing with the  chain. Thus without loss of generality,
the chain interacts with the probe  locally through the mode $q_1$, via
\be \label{Hint}
H_{\inter} =  f(\hat q_1)\hat F
\ee
with $f(x)$ expressed in the expansion form as
\be \label{fint}
f(x)=\sum_{l}\alpha_l x^l
\ee
and
\be \label{Fbbsum}
\hat F=\sum_j \frac{c_j}{\sqrt{2}}(\hat b^{\dg}_j+\hat b_j).
\ee
Additionally, there exists an onsite energy modification at site-$1$ where the probe contacted,
\be \label{Hmodf}
H'=\Delta \hat q_1^2.
\ee
Due to the periodicity considered, the above $\hat q_1$ can be replaced with any $\hat q_i$, with no difference.

The diagonalization of $H_{\chain}$
[\Eq{Hchain}] in momentum space is outlined as follows, cf.\ Ref.\,\onlinecite{Cal12}. 
For more convenience and clearance, we use the  notation in Ref.\,\onlinecite{Li02}.
Introduce
\bsube
\be
    \hat q_u = \frac{1}{\sqrt{N}}\sum_k \hat q(k)e^{iuka},
\ee
and inversely,
\be 
    \hat q(k) = \frac{1}{\sqrt{N}}\sum_u \hat q_u e^{-iuka},
\ee
\esube
where $a$ is the lattice constant
and 
$k = \frac{2n\pi}{Na} $ with $n = -\frac{N}{2}\!+\!1, \cdots , \frac{N}{2}$.
Accordingly, 
\bsube
\be
    \hat p_u = \frac{1}{\sqrt{N}}\sum_k \hat p(k)e^{-iuka} ,
\ee
and 
\be 
    \hat p(k) = \frac{1}{\sqrt{N}}\sum_u \hat p_u e^{iuka}.
\ee
\esube
Note that
$\hat q(-k) = \hat q^{\dg}(k)
$ and $\hat p(-k) = \hat p^{\dg}(k)
$. The commutation relation reads
\be
    [\hat q(k),\hat p(k')] = i\delta_{k,k'}.
\ee
Then the chain Hamiltonian $H_{\chain}$ of \Eq{Hchain} can be recast as
\begin{align}
H_{\chain}&=
\frac{1}{2m}\sum_{k}\hat p(k)\hat p^{\dg}(k) + \frac{m}{2}\sum_{k}\omega_k^2 \hat q(k)\hat q^{\dg}(k)
\nl &
= \sum_k \omega_k\big(\hat a^{\dagger}_k \hat a_k + \frac{1}{2}\big), 
\end{align}
where $\omega_k = 2\sqrt{\kappa/m}\,|{\rm sin}\frac{1}{2}ka|$. Here, we express the coordinate and momentum by creation and annihilation operators: 
\bsube\label{qpksum}
\begin{align}
\hat q(k)&=\frac{1}{\sqrt{2m\w_k}}(\hat a_{k}+\hat a_{-k}^{\dg}),
\\
\hat p(k)&=i\sqrt{\frac{m\omega_k}{2}}(\hat a_k^{\dg}-\hat a_{-k}).%
\end{align} 
\esube
From now on, we will no longer include the so-called acoustic phonon with $\w_k=0$ since it corresponds to the rigid-body translation of the entire chain. 

Fundamental statistical properties are described by correlation functions  satisfying the fluctuation--dissipation theorem (FDT) as 
\be \label{wyeq12}
\la \hat q^{\chain}_1(t)\hat q_1\ra_{\chain} =\frac{1}{\pi}\int_{-\infty}^{\infty}\!\!{\rm d}\w\,\frac{J_{\chain}(\w)}{1-e^{-\beta_1 \w}} e^{-i\w t}.
\ee
Here, $\hat q^{\chain}_1(t)\equiv e^{iH_{\chain}t}\hat q_1e^{-iH_{\chain}t}$ 
and the average is taken over the ensemble proportional to $e^{-\beta_1 H_{\chain}}$.
The spectral density can be obtained as
\be
J_{\chain}(\w)=\begin{cases}
   \frac{1}{m\w \sqrt{4\Omega_{\chain}^2-\w^2}}; & 0<|\w|< 2\Omega_{\chain}, \\
    0 ;  & |\w| \geq 2\Omega_{\chain},
\end{cases}
\label{sdJcha}
\ee
where $\Omega_{\chain}\equiv\sqrt{\kappa/m}$. More details are give in \App{appa}.
Mathematically it is not an analytic function. But physically, the molecular chain is embedded in an environment and we will apply smooth functions in later numerical simulation.

The spectral density according to the dimensionless probe operator $\hat F$ reads $J_{\probe}(\w>0)=\frac{\pi}{2}\sum_j c_j^2 \delta(\w-\epsilon_j)=-J_{\probe}(-\w)$, related to the correlation function via FDT as
\be \label{Fprocorr}
\la \hat F^{\probe}(t)\hat F\ra_{\probe} =\frac{1}{\pi}\int_{-\infty}^{\infty}\!\!{\rm d}\w\,\frac{J_{\probe}(\w)}{1-e^{-\beta_0\w}} e^{-i\w t},
\ee
with $\hat F^{\probe}(t)\equiv e^{iH_{\probe}t}\hat F e^{-iH_{\probe}t}$. The average here is taken over the ensemble proportional to $e^{-\beta_0 H_{\probe}}$. Note that the probe is at a temperature different from the chain. 
In the  numerical simulation later,
the $J_{\probe}(\w)$ will assume the Brownian oscillator form 
\be \label{probe_spe}
J_{\probe}(\w)=\frac{  \eta \Omega_{\probe} \zeta\w}{(\w^2-\Omega_{\probe}^2)^2+\zeta^2\w^2}.
\ee
The parameters $\Omega_{\probe}$ and $\zeta$ denote the characteristic frequency and the friction coefficient, respectively. The dimensionless parameter $\eta$ represents the coupling strength.

\subsection{DEOM formalism}
\label{theob}

To construct the DEOM, the  correlation functions
are expanded in terms of exponential series, i.e.,
\bsube\label{Ctexpan}
\begin{align}
 &c_0(t)\equiv\la \hat F^{\probe}(t)\hat F\ra_{\probe}\simeq\sum_{k=1}^{N_0}\eta_{0k}e^{-\gamma_{0k}t},
\\
  &c_1(t)\equiv\la \hat q^{\chain}_1(t)\hat q_1\ra_{\chain} \simeq\sum^{N_1}_{k=1}\eta_{1 k}e^{-\gamma_{1k}t},
\end{align}
\esube
via the contour integral of the Fourier integrand in FDT with the sum-over-pole decomposition.\cite{Hu10101106,Din12224103} 
To improve the numerical efficiency, we have also proposed a
time--domain Prony fitting scheme.\cite{Che22221102} 
The time-reversal relation reads
\bsube\label{FBt_corr}
\begin{align}
 &c_0^\ast(t)=\la \hat F\hat F^{\probe}(t)\ra_{\probe}\simeq\sum_{k=1}^{N_0}\eta^{\ast}_{0\bar k}e^{-\gamma_{0k}t},
\\
  &c_1^\ast(t)=\la \hat q_1\hat q^{\chain}_1(t)\ra_{\chain} \simeq\sum^{N_1}_{k=1}\eta^{\ast}_{1 \bar k}e^{-\gamma_{1k}t}.
\end{align}
\esube
The exponents $\{\gamma_{0k}\}$ and $\{\gamma_{1k}\}$ in \Eqs{Ctexpan} and (\ref{FBt_corr})  
must be either real or complex conjugate paired, and $\bar k$ is determined in the index set $ \{k=1,2,...,N_\alpha\}$
by $\gamma_{\alpha\bar k}=\gamma_{\alpha k}^{\ast}$.

The DEOM formalism starts from the dissipaton decomposition
\be \label{patonF}
\hat F=\sum_{k=1}^{N_{0}}\hat f_{0k},
\ee
and
\be \label{patonq}
\hat q_1=\sum_{k=1}^{N_{1}}\hat f_{1k},
\ee
which satisfies
\bsube
\begin{align}
\la \hat{f}^{\probe}_{0 k}(t)\hat{f}_{0k'}\ra_{\probe}=\delta_{k k'}\eta_{0k} e^{-\gamma_{0k}t},
\\ \la \hat{f}_{0 k'}\hat{f}^{\probe}_{0k}(t)\ra_{\probe}=\delta_{k k'}\eta_{0\bar k}^{\ast} e^{-\gamma_{0k}t},
\end{align}
\esube
and
\bsube
\begin{align}
\la \hat{f}^{\chain}_{1 k}(t)\hat{f}_{1k'}\ra_{\chain}=\delta_{k k'}\eta_{1k} e^{-\gamma_{1k}t},
\\ \la \hat{f}_{1 k'}\hat{f}^{\chain}_{1k}(t)\ra_{\chain}=\delta_{k k'}\eta_{1\bar k}^{\ast} e^{-\gamma_{1k}t}.
\end{align}
\esube
We then introduce the c-number-valued dissipaton moments as
\be 
M_{{\bf m}{\bf n}}(t)\equiv {\rm Tr}\bigg[\Big(\prod_{k=1}^{N_0} \hat f_{0 k}^{m_{ k}}\Big)^{\circ}\Big(\prod_ {k=1}^{N_1} \hat f_{1 k}^{n_{ k}}\Big)^{\circ}\rho_{\T}(t)\bigg],
\ee
where ${\bf m}\equiv\{m_k\geq0;k=1,\cdots,N_0\}$, ${\bf n}\equiv\{n_k\geq0;k=1,\cdots,N_1\}$, and $\rho_{\T}(t)$ obeys 
\be 
\dot \rho_{\T}(t)=-i[H_{\T},\rho_{\T}(t)].
\ee
According to the generalized diffusion equation,\cite{Yan14054105,Wan22170901}  we have
\begin{align}\label{gendiff_1}
&{\rm Tr}\Big[\Big(\prod_ {k} \hat f_{0 k}^{m_{ k}}\Big)^{\circ}\Big(\prod_ { k} \hat f_{1 k}^{n_{ k}}\Big)^{\circ}(-i)[H_{\probe},\rho_{\T}]\Big]
\nl &
= -\bigg(\sum_{k=1}^{N_0} m_k \gamma_{0k}\bigg)M_{{\bf m}{\bf n}},
\end{align}
and
\begin{align}\label{gendiff_2}
&{\rm Tr}\Big[\Big(\prod_ {k} \hat f_{0 k}^{m_{ k}}\Big)^{\circ}\Big(\prod_ { k} \hat f_{1 k}^{n_{ k}}\Big)^{\circ}(-i)[H_{\chain},\rho_{\T}]\Big]
\nl &
= -\bigg(\sum_{k=1}^{N_1} n_{k} \gamma_{1k}\bigg)M_{{\bf m}{\bf n}}.
\end{align}
The contribution of $H_{\inter}$ is
\begin{align}\label{long}
&\quad{\rm Tr}\Big[\Big(\prod_ {k} \hat f_{0 k}^{m_{ k}}\Big)^{\circ}\Big(\prod_ { k} \hat f_{1 k}^{n_{ k}}\Big)^{\circ}(-i)[H_{\inter},\rho_{\T}]\Big]
\nl &
=
-i{\rm Tr}\Big[\Big(\prod_ {k} \hat f_{0 k}^{m_{ k}}\Big)^{\circ}\Big(\prod_ { k} \hat f_{1 k}^{n_{ k}}\Big)^{\circ}f(\hat q_1)\hat F\rho_{\T}\Big]
\nl & \quad 
+i{\rm Tr}\Big[\Big(\prod_ {k} \hat f_{0 k}^{m_{ k}}\Big)^{\circ}\Big(\prod_ { k} \hat f_{1 k}^{n_{ k}}\Big)^{\circ}\rho_{\T}f(\hat q_1)\hat F\Big]
\nl &
=
-i\sum_{k'=1}^{N_0}\bigg\{ {\rm Tr}\Big[\Big(\prod_ {k} \hat f_{0 k}^{m_{ k}+\delta_{k{k'}}}\Big)^{\circ}\Big(\prod_ { k} \hat f_{1 k}^{n_{ k}}\Big)^{\circ}f(\hat q_1)\rho_{\T}\Big]
\nl & \quad
+m_{k'} \eta_{0{k'}}{\rm Tr}\Big[\Big(\prod_ {k} \hat f_{0 k}^{m_{ k}-\delta_{k{k'}}}\Big)^{\circ}\Big(\prod_ { k} \hat f_{1 k}^{n_{ k}}\Big)^{\circ}f(\hat q_1)\rho_{\T}\Big]\bigg\}
\nl & \quad
+i\sum_{k'=1}^{N_0}\bigg\{ {\rm Tr}\Big[\Big(\prod_ {k} \hat f_{0 k}^{m_{ k}+\delta_{k{k'}}}\Big)^{\circ}\Big(\prod_ { k} \hat f_{1 k}^{n_{ k}}\Big)^{\circ}\rho_{\T}f(\hat q_1)\Big]
\nl & \quad
+m_{k'} \eta_{0\bar {k'}}^{\ast}{\rm Tr}\Big[\Big(\prod_ {k} \hat f_{0 k}^{m_{ k}-\delta_{k{k'}}}\Big)^{\circ}\Big(\prod_ { k} \hat f_{1 k}^{n_{ k}}\Big)^{\circ}\rho_{\T}f(\hat q_1)\Big]\bigg\}
\nl &
\equiv 
-i\sum_{k'=1}^{N_0} \Big\{M_{{\bf m}_{{k'}}^{+}{\bf n}}\big(t;[f(\hat q_1)]^{>}\big)-M_{{\bf m}_{{k'}}^{+}{\bf n}}\big(t;[f(\hat q_1)]^{<}\big)
\nl & \quad\quad\quad\quad\quad\,
+m_{k'}\eta_{0{k'}} M_{{\bf m}_{{k'}}^{-}{\bf n}}\big(t;[f(\hat q_1)]^{>}\big)
\nl & \quad\quad\quad\quad\quad\,
-m_{k'} \eta_{0\bar {k'}}^{\ast}M_{{\bf m}_{k'}^{-}{\bf n}}\big(t;[f(\hat q_1)]^{<}\big)
\Big\}.
\end{align}
In the second identity, the generalized Wick's theorem is used.\cite{Yan14054105,Wan22170901} Similarly, the contribution of $H'$ is
\begin{align}
&\quad{\rm Tr}\Big[\Big(\prod_ {k} \hat f_{0 k}^{m_{ k}}\Big)^{\circ}\Big(\prod_ { k} \hat f_{1 k}^{n_{ k}}\Big)^{\circ}(-i)[H',\rho_{\T}]\Big]
\nl &
=
-i\Delta{\rm Tr}\Big[\Big(\prod_ {k} \hat f_{0 k}^{m_{ k}}\Big)^{\circ}\Big(\prod_ { k} \hat f_{1 k}^{n_{ k}}\Big)^{\circ}\hat q_1^2\rho_{\T}\Big]
\nl & \quad 
+i\Delta{\rm Tr}\Big[\Big(\prod_ {k} \hat f_{0 k}^{m_{ k}}\Big)^{\circ}\Big(\prod_ { k} \hat f_{1 k}^{n_{ k}}\Big)^{\circ}\rho_{\T}\hat q_1^2\Big]
\nl &
=
-i\Delta\Big[M_{{\bf m}{\bf n}}\big(t;(\hat q_1^2)^{>}\big)-M_{{\bf m}{\bf n}}\big(t;(\hat q_1^2)^{<}\big)\Big].
\end{align}

As a result, \Eqs{gendiff_1}--(\ref{long}) give rise to
\begin{align} \label{dm_eom}
\frac{\rm d}{{\rm d}t}M_{{\bf m}{\bf n}}(t)&=-\bigg(\sum_{k=1}^{N_0} m_k \gamma_{0k}+\sum_{k=1}^{N_1} n_{k} \gamma_{1k}\bigg)M_{{\bf m}{\bf n}}(t)
\nl &\hspace{-2em}
-i\Delta \Big[M_{{\bf m}
{\bf n}^{(2)}
{\bf \ti n}^{(2)}}^{>}(t)-M_{{\bf m}
{\bf n}^{(2)}
{\bf \ti n}^{(2)}}^{<}(t)\Big]
\nl & \hspace{-2em}
-i\sum_{k=1}^{N_0} \Big (\wti M_{{\bf m}_{k}^{+}{\bf n}}^{>}(t)-\wti M_{{\bf m}_{k}^{+}{\bf n}}^{<}(t)\Big)
\nl & \hspace{-2em}
-i\sum_{k=1}^{N_0} m_k\Big(\eta_{0k} \wti M_{{\bf m}_{k}^{-}{\bf n}}^{>}(t)
- \eta_{0\bar k}^{\ast}\wti M_{{\bf m}_{k}^{-}{\bf n}}^{<}(t)
\Big)
\end{align}
with [cf.\,\Eq{fint}]
\begin{align}\label{expansion}
\wti M_{{\bf m}{\bf n}}^{\gler}(t)&\equiv M_{{\bf m}{\bf n}}\big(t;[f(\hat q_1)]^{\gler}\big)
\nl &
=\sum_l \alpha_l M_{{\bf m}{\bf n}}\big(t;(\hat q_1^{l})^{\gler}\big)
\nl &\equiv\sum_l \alpha_l M_{{\bf m}
{\bf n}^{(l)}
{\bf \ti n}^{(l)}}^{\gler}(t).
\end{align}
In the last identity, ${\bf n}^{(l)}{\bf \ti n}^{(l)}$ denotes the actual index set for $\hat q_1$--related dissipatons,
with numbers in  
${\bf n}^{(l)}$
for the original $\bf n$--dissipatons left and
numbers in 
${\bf \ti n}^{(l)}$ for the newly added dissipatons, after the operation of $(\hat q_1^{l})^{\gler}$.
Evidently $M_{{\bf m}{\bf n}}\big(t;(\hat q_1^{l=0})^{\gler}\big)=M_{{\bf m}{\bf n}}(t)$.
Recursively, 
\begin{align}\label{rec}
    M_{{\bf m}{{\bf n}^{(l)}
{\bf \ti n}^{(l)}}}
    \big(t;&(\hat q_1^{l+1})^{\gler}\big)
    =\sum_{k=1}^{N_1} \Big[M_{{\bf m}{{\bf n}^{(l)}
{\bf \ti n}^{(l)+}_{k}}}\big(t;(\hat q_1^{l})^{\gler}\big)
    \nl &+n_k^{(l)} \eta_{1k}^{\gler} M_{{\bf m}{{\bf n}^{(l)-}_{k}
{\bf \ti n}^{(l)}}}\big(t;(\hat q_1^{l})^{\gler}\big)
    \nl &+\ti n_k^{(l)} \bar\eta_{1k}^{\gler} M_{{\bf m}{{\bf n}^{(l)}
{\bf \ti n}^{(l)-}_{k}}}\big(t;(\hat q_1^{l})^{\gler}\big)
    \Big]
\end{align}
with $\eta^{\greater}_{1k}=\eta_{1k}$, $\eta^{\lesser}_{1 k}=\eta^{\ast}_{1\bar k}$, and
$\bar\eta^{\gler}_{1k}=(\eta_{1k}+\eta^{\ast}_{1\bar k})/2$.\cite{Xu18114103}
This is the key step in treating higher-order couplings. 
Equation (\ref{dm_eom}), together with \Eqs{expansion} and (\ref{rec}), 
constitute our final formalism of the c-number-valued DEOM for dissipaton moments.
The DEOM for density matrices with higher-order bath couplings have also been constructed in a similar manner.\cite{Zhu2667}
Apparently the coupling on the part of $\hat F$ will be easily treated if it is nonlinear.
The formalism can also be extended straightforwardly to cases where multiple sites are influenced by the probe.

To conclude this section, we introduce the heat current between the probe and the chain that will be simulated later in \Sec{secnum}. Define
\be 
\hat I\equiv -\frac{{\rm d}H_{\probe}}{{\rm d}t}=-i[H_{\inter},H_{\probe}]=f(\hat q_1)\hat\Phi,
\ee
with 
$ 
\hat \Phi\equiv i[H_{\probe},\hat F]=\sum_{k=1}^{N_0} \hat \varphi_{0k}$ which is of the dimension of energy as $\hat F$ is selected to be dimensionless.
The expectation of $\hat I$ can be evaluated via the dissipaton’s momentum algebra as \cite{Wan20041102} [cf.\,\Eq{fint} and \Eq{rec}]
\begin{align}
  I(t) &= {\rm Tr}\big[
f(\hat q_1)\hat \Phi\rho_{\T}(t)\big]
 = \sum_k  {\rm Tr}\big[
f(\hat q_1)\hat \varphi_{0k}\rho_{\T}(t)\big]
\nl &
=-\sum_{kl} \alpha_l \gamma_{0k}M_{{\bf 0}_{k}^{+}{\bf 0}}\big(t;(\hat q_1^{l})^{>}\big).
\end{align}

The 1D chain exemplified in this work can be readily generalized to 2D or 3D harmonic or noninteracting particle systems.
Moreover, anharmonic or strongly correlated systems may also be applied, as far as their spectral densities can be achieved.
Finally, we note that the method presented in this subsection can be straightforwardly analogized to electronic transport problems. 

\begin{figure}[t]
\includegraphics[width=0.9\columnwidth]{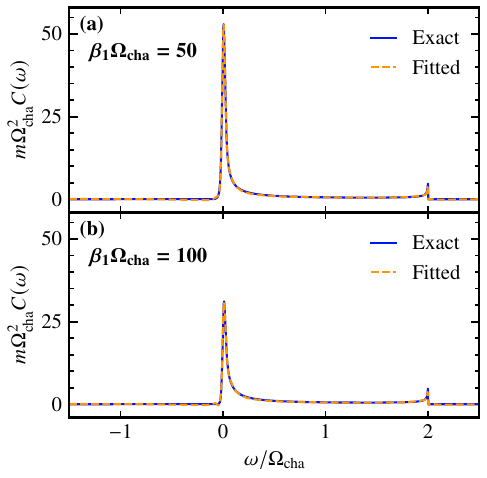}
\caption{The numerically fitted   chain spectra $C(\w)$ at different temperatures, compared with the exact ones, $J_{\chain}(\w)/(1-e^{-\beta_1\w})$ [cf.\,\Eqs{wyeq12} and (\ref{sdJcha})].
}\label{fig2}
\end{figure}

\section{Numerical demonstration}
\label{secnum}
For the numerical demonstrations, we set $\Omega_{\chain}=1$ as the unit. We keep the temperature ratio between the probe and the chain to be $T_{0}/T_{1}=50$, and exemplify with $T_{0}=1$ and $0.5$. Figure \ref{fig2} shows the numerically fitted chain correlation via the time-domain Prony fitting scheme,\cite{Che22221102} depicted in the frequency domain at the two temperatures, $T_{1}=0.02$ and $0.01$, in comparison to the exact ones, $C(\w)\equiv J_{\chain}(\w)/(1-e^{-\beta_1\w})$ [cf.\,\Eqs{wyeq12} and (\ref{sdJcha})]. 
In practice, the chain spectral density $J_{\chain}(\omega)$ is smoothed according to
\[
J_{\chain}(\omega)\rightarrow J_{\chain}(\omega)
\Big(1-e^{-\frac{(\omega-2\Omega_{\chain})^2}{2\sigma_1^2}}\Big)
\Big(1-e^{-\frac{\omega^2}{2\sigma_0^2}}\Big),
\]
with $\sigma_0=0.01$ and $\sigma_1=0.001$, preserving the relative peak heights.
For the probe adopting the Brownian oscillator model, \Eq{probe_spe}, its correlation function, \Eq{Fprocorr}, is expanded using the optimized Pad\'{e} scheme in Ref.\,\onlinecite{Din12224103}. 
In the following, we set $m=1$ for simplicity. Assuming that \(H_{\inter}\) has even parity and noting that \(F\) is odd, we set \(\alpha_0=\alpha_2=0\).
In all numerical demonstrations, we choose $\alpha_1=0.1$. In our calculations, we set $\alpha_3< 0$ when higher-order effects are explored. We find that, compared with positive $\alpha_3$, a negative value of $\alpha_3$ renders the composite system more stable.
We conjecture that this is related to whether the total Hamiltonian is lower-bounded.

\begin{figure}[t]
\includegraphics[width=0.9\columnwidth]{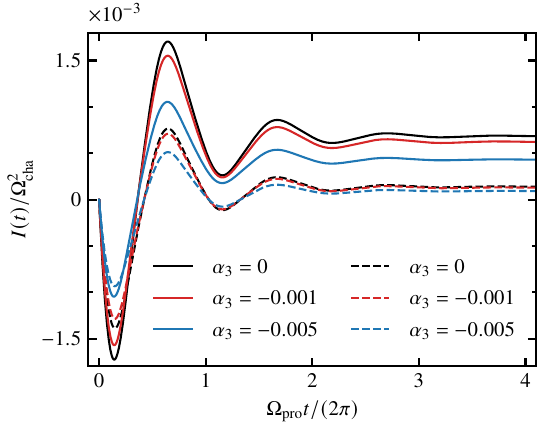}
\caption{Transient currents at different temperatures (solid curves: $T_0=1$; dashed curves: $T_0=0.5$). We choose $\alpha_3= 0$, $-0.001$, and $-0.005$ for the higher-order coupling [cf.\,\Eqs{Hint} and (\ref{fint})].
}\label{fig3}
\end{figure}

\begin{figure}[t]
\includegraphics[width=0.9\columnwidth]{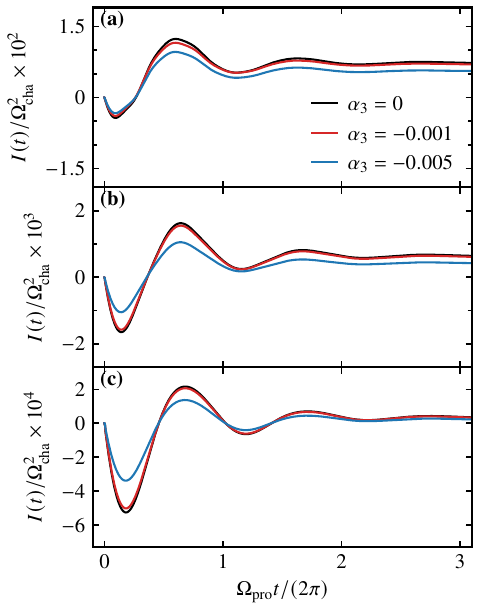}
\caption{Transient currents with $T_0=1$ and $\Delta=0$, evaluated by varying the probe spectral density parameters: (a) $\Omega_{\probe} = 0.5, \zeta = 0.25, \eta = 1$, (b) $\Omega_{\probe} = 1, \zeta = 0.5,  \eta = 0.25$, and (c) $\Omega_{\probe} = 2, \zeta = 1, \eta = 0.0625$, for comparison.
}\label{fig4}
\end{figure}

\begin{figure}[t]
\includegraphics[width=0.9\columnwidth]{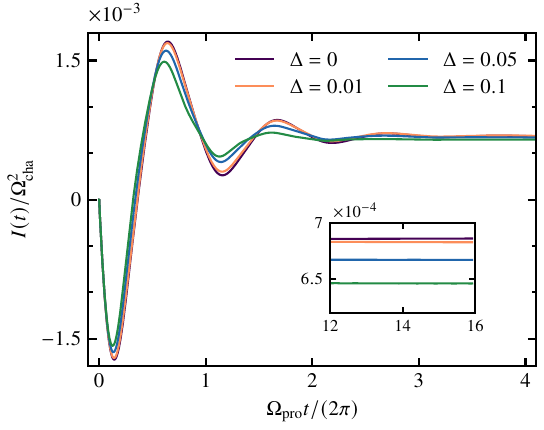}
\caption{Transient currents with $T_0=1$ and $\alpha_3=0$, evaluated by varying the onsite energy modification $\Delta$. Other parameters related to the probe are the same as in \Fig{fig3}. Shown in the inset are asymptotic behaviors of the calculated currents.
}\label{fig5}
\end{figure}

Figure \ref{fig3} exhibits the evaluated heat currents from probe to chain, at the two temperatures, with $\Delta=0$, and $\Omega_{\probe} = 1, \zeta = 0.5$ and $\eta = 0.25$ in \Eq{probe_spe}.
As expected, the heat current increases with the temperature bias, since a larger temperature difference provides a stronger thermodynamic driving force. In contrast, the nonlinear effect significantly suppresses the heat current. This is also observed in \Fig{fig4} which exhibits the currents with $T_0=1$ and $\Delta=0$, evaluated by varying the probe spectral density parameters.  
This reduction may arise from the fact that higher-order interactions, i.e.\,anharmonicity, introduces additional phonon–phonon scattering and inelastic transport channels, which weaken coherent energy transfer and lower the effective transmission efficiency. Moreover, it is observed the influence becomes more pronounced at high temperatures than at low temperatures.

Figure \ref{fig5} presents the calculated currents under different onsite energy modifications.
As the onsite energy modification $\Delta$ increases, we observe a corresponding increase in the oscillation frequency of the current measured by the probe. Moreover, the steady-state heat current decreases as $\Delta$ increases.
This may be attributed to the fact that larger onsite energy gap suppresses the heat inflow.

\section{Summary}
\label{secsum}
To summarize, we have developed an efficient approach based on  dissipaton-equation-of-motion (DEOM) method to study the local thermal properties of a one-dimensional molecular chain  coupled to a probe. The coupling involves higher-order interactions, i.e.\,anharmonicity, between the local site of the chain and the probe, and an onsite energy modification induced by the probe.
Starting from the Hamiltonian of the chain--probe composite and the associated spectral densities, we derive the  c-number-valued DEOM for dissipaton moments which is efficient meanwhile fully nonperturbative and non-Markovian and evaluate the heat current between the probe and the chain within a fully quantum-mechanical framework. 

The main advance of this work is that the formulation of the DEOM for dissipaton moments is expressed in terms of ordinary c-number variables, rather than density matrices, in the coupled equations of motion. This formulation is achieved without introducing additional approximations or restrictions, while offering significant computational efficiency. For bilinear chain--probe coupling, the DEOM approach is shown to be consistent with the analytic results from fundamental quantum mechanics. Furthermore, the dissipaton algebra allows higher-order chain--probe interactions to be incorporated naturally, leading to iterative cross-tier connections in the resulting hierarchically coupled equations. This demonstrates the flexibility of the present framework in treating anharmonic and many-body effects beyond the bilinear regime.

The numerical results reveal the effects of temperature,  frequency, onsite energy modification, and higher-order interactions in determining the heat transport behavior of the composite system. These results show that a local probe can serve as an effective tool for characterizing the thermal response of low-dimensional systems and for investigating the microscopic mechanisms governing local energy transfer.
From a broader perspective, the present method can be naturally extended to higher-dimensional materials and to more intricate scenarios involving structured multiple-site probe-chain couplings, strong correlations, and nonequilibrium many-body effects. Its analogue to electronic transport problems may also provide valuable insight into the roles of local probing, dissipation, and non-Markovianity in determining transport behaviors. We therefore anticipate that this approach will offer a versatile platform for future studies of quantum energy and charge transport in complex molecular and condensed-matter systems.

\section*{Acknowledgment}

Support from the National Natural Science Foundation of China (Grant Nos.\   
22373091 and 22573099) is gratefully acknowledged. 
Simulations were performed on the robotic AI-Scientist platform of Chinese Academy of Sciences.
The authors are indebted to Yu Su and Zi-Hao Chen for invaluable and constructive discussions and software support.

\appendix
\section{Detailed derivation for \Eq{sdJcha}}
\label{appa}

In this appendix, we present how to obtain the spectral density in \Eq{sdJcha}.
We will derive the general ones, $\la \hat q^{\chain}_u(t)\hat q_v\ra_{\chain}$ and $J^{\chain}_{uv}(\w)$.
For $\hat q^{\chain}_u(t)$ defined similarly as $\hat q^{\chain}_1(t)$ below \Eq{wyeq12}, we have
\be
    \hat q^{\chain}_u(t)
=
\frac{1}{\sqrt{N}}
\sum_{k}
\frac{1}{\sqrt{2m\omega_k}}\,
e^{iuka}
[
\hat a^{\chain}_k(t)
+
\hat a^{\dg{\chain}}_{-k}(t)
]
\ee
where $\hat a^{\chain}_k(t)=e^{-i\w_k t}\hat a_k$ and $\hat a^{\dg{\chain}}_{-k}(t)=e^{i\w_k t}\hat a^{\dg}_{-k}$.
As a result,
\begin{align}\label{app11}
\la \hat q^{\chain}_u(t) \hat q_v\ra_{\chain} 
&=\frac{1}{2mN}\sum_{k} \frac{1 
}{\w_k}
[\bar n_{k}(\beta_1)e^{i\w_k t}e^{-i(u-v)ka}
\nl & \qquad
+(\bar n_{k}(\beta_1)+1)e^{-i\w_k t}e^{i(u-v)ka}],
\end{align}
where $\bar n_{k}(\beta_1)=1/(e^{\beta_1\w_k}-1)$. One may recast \Eq{app11} as
\begin{align}
\la \hat q^{\chain}_u(t)\hat q_v\ra_{\chain}=
\frac{1}{\pi}\int_{-\infty}^{\infty}\!\!{\rm d}\w\,\frac{J^{\chain}_{uv}(\w)}{1-e^{-\beta_1 \w}} e^{-i\w t},
\end{align}
with the spectral density
\begin{align}
J^{\chain}_{uv}(\w>0)&=\frac{\pi/N}{2m\w}\sum_ke^{i(u-v)ka}\,\delta(\w-\w_k)
\nl &=-J^{\chain}_{vu}(-\w).
\end{align}
It can be further evaluated by replacing 
\be\nonumber
\sum_k\Rightarrow N\frac{a}{2\pi}\lim_{s\rightarrow\frac{\pi}{a}}\int_{-s}^{s}{\rm d}k
\ee
which gives rise to
\begin{align}
    J^{\chain}_{uv}(\w>0)=\frac{a}{2m\w} \int_0^{\frac{\pi}{a}}\!\! {\rm d}k\cos[(u-v)ka]\delta(\w-\w_k).
\end{align}
Noting that $\omega_k = 2\sqrt{\kappa/m}\,|{\rm sin}\frac{1}{2}ka|$ and $\Omega_{\chain}=\sqrt{\kappa/m}$, we obtain for $0<|\w|< 2\Omega_{\chain}$
\begin{align}
J^{\chain}_{uv}(\w)&=  \frac{\cos(2(u-v){\rm arcsin}\frac{\w}{2\Omega_{\chain}})}{m\w \sqrt{4\Omega_{\chain}^2-\w^2}}
\nl 
&= \frac{T_{2(u-v)}\big(\sqrt{1-\frac{\w^2}{4\Omega^2_{\chain}}}\,\big)}
{m\w \sqrt{4\Omega_{\chain}^2-\w^2}}\,,
\end{align}
where $T_n(x)$ is the Chebyshev polynomial of the first kind.
Equation (\ref{sdJcha}) is resulted with $u=v=1$.

\section{Condition of bi-linear coupling}
\label{appb}

In this appendix, we consider the bi-linear case, with $f(\hat q_1)=\alpha_1\hat q_1$ in \Eq{Hint} and $\Delta=0$ in \Eq{Hmodf}, that makes the total composite behave Gaussian and \Eqs{dm_eom}--(\ref{rec}) become closed equations for the first and second dissipaton moments. In this bi-linear case, we can also derive,
from fundamental quantum mechanics, the first and second moments of
$\hat q_1$ and $\hat F$. We will show all these results are consistent.

\subsection{DEOM approach}
\label{appB1}
Equations (\ref{expansion}) and (\ref{rec}) 
for $f(\hat q_1)=\alpha_1\hat q_1$
reduce to
\begin{align}\label{appb1}
\wti M_{{\bf m}{\bf n}}^{\gler}(t)&= \alpha_1 M_{{\bf m}{\bf n}}\big(t;\hat q_1^{\gler}\big)
\nl & =\alpha_1
\sum_k \big[M_{{\bf m}{\bf n}_{k}^{+}}(t)+n_k \eta_{1k}^{\gler} M_{{\bf m}{\bf n}_{k}^{-}}(t)\big]
.
\end{align}
Equation (\ref{dm_eom}) then becomes
\begin{align} \label{appb2}
\frac{\rm d}{{\rm d}t}M_{{\bf m}{\bf n}}(t)&=-\Big(\sum_k m_k \gamma_{0k}+\sum_{k} n_{k} \gamma_{1k}\Big)M_{{\bf m}{\bf n}}(t)
\nl & \hspace{-2.5em}
-i\alpha_1\sum_k\sum_{k'} n_{k'} (\eta_{1k'}-\eta^\ast_{1\bar k'}) M_{{\bf m}_{k}^{+}{\bf n}_{k'}^{-}}(t)
\nl & \hspace{-2.5em}
-i\alpha_1\sum_k\sum_{k'} m_k(\eta_{0k}-\eta_{0\bar k}^{\ast})M_{{\bf m}_{k}^{-}{{\bf n}}^{+}_{k'}}(t)
\nl & \hspace{-2.5em}
-i\alpha_1\sum_k\sum_{k'} m_kn_{k'} (\eta_{0k}\eta_{1k'}
- \eta_{0\bar k}^{\ast}\eta^\ast_{1\bar k'})M_{{\bf m}_{k}^{-}{\bf n}^{-}_{k'}}(t).
\end{align} 
Denote for $r=0,1$
\be\label{faverage}
    f_{rk}(t) \equiv {\rm Tr}[(\hat f_{r k})^{\circ}\rho_{\T}(t)] 
\ee
and
\be
    \sigma_{r k, r' k'}(t) \equiv {\rm Tr}[(\hat f_{r k} \hat f_{r' k'})^{\circ}\rho_{\T}(t)]=    \sigma_{r' k', r k}.
\ee
From \Eq{appb2}, they obey 
\bsube
\begin{align}\label{appb5a}
      \dot f_{0 k}(t) &=-\gamma_{0 k}f_{0 k}(t) -i\alpha_1 (\eta_{0k}-\eta_{0\bar k}^{\ast}) \sum_{k'}f_{1k'}(t),\\
      \dot f_{1 k}(t) &= -\gamma_{1 k}f_{1 k}(t) 
      -i\alpha_1
      (\eta_{1k}-\eta^\ast_{1\bar k})  \sum_{k'}f_{0 k'}(t),
\label{appb5b}
\end{align}
\label{appb5}
\esube
and
\bsube
\begin{align} \label{b6a}
  \dot \sigma_{\tL k,\tL k'} &= -(\gamma_{\tL k} + \gamma_{\tL k'})\sigma_{\tL k,\tL k'}  -i\alpha_1 (\eta_{0k}-\eta_{0\bar k}^{\ast}) \sum_{j}\sigma_{\tL k',\tR j}
  \nl  & \quad\,
   -i\alpha_1 (\eta_{0k'}-\eta_{0\bar k'}^{\ast})  \sum_{j}\sigma_{\tL k,\tR j}\,, 
  \\ \label{b6b}
  \dot \sigma_{\tL k,\tR k'} &= -(\gamma_{\tL k} + \gamma_{\tR k'})\sigma_{\tL k,\tR k'} -i\alpha_1 (\eta_{\tL k}\eta_{\tR k'} - \eta_{\tL \bar k}^*\eta_{\tR \bar k'}^*) 
  \nl & \quad \, 
  -i\alpha_1 (\eta_{0k}-\eta_{0\bar k}^{\ast})\sum_{j}\sigma_{\tR k',\tR j}
    \nl & \quad \,
  -i\alpha_1 (\eta_{1k'}-\eta_{1\bar k'}^{\ast})\sum_{j}\sigma_{\tL k,\tL j}\,,\\ 
  \label{b6c}
  \dot \sigma_{\tR k,\tR k'} &= -(\gamma_{\tR k} + \gamma_{\tR k'})\sigma_{\tR k,\tR k'}  -i\alpha_1
      (\eta_{1k}-\eta^\ast_{1\bar k})\sum_{j}\sigma_{\tL j,\tR k'}
\nl &\quad\, 
   -i\alpha_1
      (\eta_{1k'}-\eta^\ast_{1\bar k'})\sum_{j}\sigma_{\tL j,\tR k}\,.
\end{align}
\label{appb6}
\esube

\subsection{Fundamental quantum mechanics approaches}
\label{appB2}
The total Hamiltonian is now recast as
\begin{align}
H_{\T}&= H_{\probe} + H_{\chain} + \alpha_1 \hat q_1 \hat F
\nl &=\sum_k \omega_k\hat a^{\dagger}_k \hat a_k +\sum_{j}\epsilon_{j} \hat b_{j}^{\dg}\hat b_j
\nl & \quad
+\alpha_1\Big[\sum_k\frac{A_k}{\sqrt{2}}(\hat a_{-k}^{\dg}+\hat a_{k})\Big]\Big[\sum_j\frac{c_j}{\sqrt{2}}(\hat b^{\dg}_j+\hat b_j)\Big]
\end{align}
with
$A_k\equiv e^{ika}/\sqrt{Nm\w_k}=A^\ast_{-k}$.
Denote the operator evolution in the Heisenberg picture $\hat O(t)\equiv e^{iH_{\T}t}\hat O
e^{-iH_{\T}t}$
and 
[cf.\ \Eqs{Fbbsum} and (\ref{qpksum})]
\bsube
\begin{align}    
 &\hat q_1(t)=\sum_k\hat q(t;k)=\sum_k\frac{A_k}{\sqrt{2}}[\hat a^\dg_{-k}(t)+\hat a_k(t)],
\\
 &\hat F(t)=\sum_j\hat F_j(t)=\sum_j\frac{c_j}{\sqrt{2}}[\hat b^\dg_{j}(t)+\hat b_j(t)].
\end{align}
\esube
We have from the Heisenberg equations
\bsube
\begin{align}
\ddot{\hat q}(t;k)&=-\w^2_k{\hat q}(t;k)
-{\alpha_1}\w_k|A_k|^2
\hat F(t),
\\
\ddot{\hat F}_j(t)&=-\epsilon^2_j{\hat F}_j(t)
-{\alpha_1}\epsilon_j c_j^2
\hat q_1(t).
\end{align}
\esube
The solutions are
\bsube
\begin{align}
{\hat q}(t;k)&=
{\hat q}^{\chain}(t;k)
-\frac{\alpha_1}{Nm\w_k}
\int^t_0\!\d\tau\,\sin[\w_k(t-\tau)]\hat F(\tau),
\\
{\hat F}_j(t)&=
{\hat F}^{\probe}_j(t)
-{\alpha_1}c_j^2
\int^t_0\!\d\tau\,
\sin[\epsilon_j(t-\tau)]\hat q_1(\tau),
\end{align}
\esube
which lead to
\bsube
\begin{align}
{\hat q}_1(t)&=
{\hat q}^{\chain}_1(t)
-i\alpha_1
\int^t_0\!\d\tau\,
\big[\la \hat q^{\chain}_1(t-\tau)\hat q_1\ra_{\chain}
\nl & \qquad\qquad
-\la \hat q_1
\hat q^{\chain}_1(t-\tau)\ra_{\chain}\big]\hat F(\tau),
\\
{\hat F}(t)&=
{\hat F}^{\probe}(t)
-i{\alpha_1}
\int^t_0\!\d\tau\,
\big[\la \hat F^{\probe}(t-\tau)\hat F\ra_{\probe}
\nl & \qquad\qquad
-\la \hat F\hat F^{\probe}(t-\tau)\ra_{\probe}\big]\hat q_1(\tau).
\end{align}
\label{qFsolution}
\esube

In this paper, moments are averaged according to the initial state, $\rho_{\T}\propto e^{-\beta_1 H_{\chain}}
\otimes
e^{-\beta_0 H_{\probe}}$, denoted as
$\la\hat O(t)\ra={\rm Tr}[\hat O(t)\rho_{\T}]={\rm Tr}[\hat O\rho_{\T}(t)]$.
For the first order moments, as $\la\hat q^{\chain}_1(t)\ra=\la{\hat F}^{\probe}(t)\ra=0$, 
we obtain from \Eq{qFsolution}
\bsube\label{qF_solution}
\begin{align}
\label{qFa}
\la {\hat q}_1(t) \ra &=
-i\alpha_1
\int^t_0\!\d\tau\,
\big[\la \hat q^{\chain}_1(t-\tau)\hat q_1\ra_{\chain}
\nl & \qquad\qquad
-\la \hat q_1
\hat q^{\chain}_1(t-\tau)\ra_{\chain}\big] \la\hat F(\tau)\ra,
\\
\la {\hat F}(t) \ra&=
-i{\alpha_1}
\int^t_0\!\d\tau\,
\big[\la \hat F^{\probe}(t-\tau)\hat F\ra_{\probe}
\nl & \qquad\qquad
-\la \hat F\hat F^{\probe}(t-\tau)\ra_{\probe}\big] \la\hat q_1(\tau)\ra.
\label{qFb}
\end{align}
\esube
To continue, for convenience, let us denote
\bsube\label{phidefappb}
\begin{align}
\phi_0(t)&\equiv i[c_0(t)-c_0^\ast(t)] \nl &=i\big(\la \hat F^{\probe}(t)\hat F\ra_{\probe}
-\la \hat F\hat F^{\probe}(t)\ra_{\probe}\big)
\nl & =
i\big(\hat F^{\probe}(t)\hat F
-\hat F\hat F^{\probe}(t)\big)
\nl & =
i\sum_k(\eta_{0k}-\eta^\ast_{0\bar k})
e^{-\gamma_{0k}t},
\\
\phi_1(t)&\equiv  i[c_1(t)-c_1^\ast(t)] \nl &=i\big(\la \hat q^{\chain}_1(t)\hat q_1\ra_{\chain}
-\la \hat q_1
\hat q^{\chain}_1(t)\ra_{\chain}\big)
\nl & =
i\big(\hat q^{\chain}_1(t)\hat q_1
-\hat q_1
\hat q^{\chain}_1(t)\big)
\nl & =
i\sum_k(\eta_{1k}-\eta^\ast_{1\bar k})
e^{-\gamma_{1k}t}.
\end{align}
\esube
By \Eqs{patonF}--(\ref{patonq}) and \Eq{faverage}, we have $\la {\hat q}_1(t) \ra=\sum_k f_{1k}(t)$
and
$\la {\hat F}(t) \ra=\sum_k f_{0k}(t)$
which, from \Eqs{qF_solution} and (\ref{phidefappb}), satisfy
\bsube
\begin{align}
f_{0k}(t)&=
-i{\alpha_1}
\int^t_0\!\d\tau\,
(\eta_{0k}-\eta^\ast_{0\bar k})
e^{-\gamma_{0k}(t-\tau)}
\sum_{k'}f_{1k'}(\tau),
\\
f_{1k}(t)&=
-i{\alpha_1}
\int^t_0\!\d\tau\,
(\eta_{1k}-\eta^\ast_{1\bar k})
e^{-\gamma_{1k}(t-\tau)}
\sum_{k'}f_{0k'}(\tau),
\end{align}
\esube
They are rightly the solutions to 
\Eqs{appb5a} and (\ref{appb5b}).

The equivalence can also be obtained with the Dyson equation,
\begin{align}\label{B31}
    e^{-i\mathcal{L}_{\T}t} = e^{ -i\mathcal{L}_{0} t} - i\alpha_1 \! \int^t_0 \! \! {\rm d}\tau \, e^{ -i \mathcal{L}_{0}(t-\tau)}  (\hat q_1\hat F)^{\times} e^{ -i\mathcal{L}_{\T} \tau} ,
\end{align}
where $H_0 \equiv H_{\probe} + H_{\chain} $, $\mathcal{L}_{\T}\cdot \equiv [H_{\T}, \cdot]$, $\mathcal{L}_0 \cdot = [H_0, \cdot]$, and $(\hat q_1 \hat F)^{\times} \cdot \equiv [\hat q_1 \hat F, \cdot]$.
We then obtain 
\begin{align}
    \la \hat q_1(t) \ra 
    &= - i\alpha_1 \! \int^t_0 \! {\rm d}\tau \, {\rm Tr}\Big\{  [\hat q_1^{\chain}(t - \tau), \hat q_1 \hat F] \rho_{\T}(\tau) \Big\} \nl
    &= - \alpha_1 \! \int^t_0\! \! {\rm d}\tau \, \phi_1(t- \tau)\la\hat F(\tau)\ra,
\end{align}
where we have used $\phi_1(t) = i[\hat q_1^{\chain}(t), \hat q_1]$. 
Similarly, 
\begin{align}
    \la \hat F(t) \ra 
    &= - i\alpha_1 \! \int^t_0 \! {\rm d}\tau \, {\rm Tr}\Big\{  [\hat F^{\probe}(t - \tau), \hat q_1 \hat F] \rho_{\T}(\tau) \Big\} \nl
    &= - \alpha_1 \! \int^t_0\! \! {\rm d}\tau \, \phi_0(t- \tau)\la\hat q_1(\tau)\ra,
\end{align}
where we have used $\phi_0(t) = i[\hat F^{\probe}(t), \hat F]$. They are just \Eqs{qFa} and (\ref{qFb}) thus consistent with \Eqs{appb5a} and (\ref{appb5b}).

%

Turn now to the second order moments. Note that $\la{\hat F}^{\probe}(t){\hat F}^{\probe}(t)\ra=\la\hat F^2\ra_{\probe}=c_0(0)$,
$\la\hat q^{\chain}_1(t)
\hat q^{\chain}_1(t)\ra=\la\hat q^2_1\ra_{\chain}=c_1(0)$ and $\la\hat q^{\chain}_1(t){\hat F}^{\probe}(t)\ra=\la{\hat F}^{\probe}(t)\hat q^{\chain}_1(t)\ra=0$.
Using \Eq{B31}, we have
\bsube\label{appbcheck1}
\begin{align}
    \la &\hat F^2(t) \ra 
    = c_0(0)  - i\alpha_1 \! \int^t_0 \! {\rm d}\tau \, {\rm Tr}\Big\{  [\hat F^{\probe\,2}(t - \tau), \hat q_1 \hat F] \rho_{\T}(\tau) \Big\} \nl
    &= c_0(0) - 2\alpha_1 \! \int^t_0\! \! {\rm d}\tau \, \phi_0(t- \tau) {\rm Tr}[\hat F^{\probe}(t- \tau) \hat q_1 \rho_{\T}(\tau)],
\end{align}
\begin{align}
    \la&\hat q_1(t) \hat F(t) \ra=\la\hat F(t) \hat q_1(t) \ra \nl
    &= -i\alpha_1 \! \int^t_0 \! \! {\rm d}\tau \, {\rm Tr}\Big\{  [\hat q^{\chain}_1(t-\tau) \hat F^{\probe}(t - \tau), \hat q_1 \hat F] \rho_{\T}(\tau) \Big\} \nl
     &= -\alpha_1 \! \int^t_0 \! \! {\rm d}\tau \, \Big( \phi_0(t-\tau) {\rm Tr}[\hat q_1^{\chain}(t-\tau) \hat q_1 \rho_{\T}(\tau)] \nl 
     &\qquad\qquad + \phi_1(t-\tau) {\rm Tr}[\hat F \hat F^{\probe}(t-\tau)\rho_{\T}(\tau)]  \Big),
\end{align}
and
\begin{align}
    \la&\hat q_1^2(t) \ra 
    = c_1(0)  - i\alpha_1 \! \int^t_0 \! {\rm d}\tau \, {\rm Tr}\Big\{  [\hat q^{\chain\,2}_1(t - \tau), \hat q_1 \hat F] \rho_{\T}(\tau) \Big\} \nl
    &= c_1(0) - 2\alpha_1 \! \int^t_0\! \! {\rm d}\tau \, \phi_1(t- \tau) {\rm Tr}[\hat q_1^{\chain}(t- \tau) \hat F \rho_{\T}(\tau)].
\end{align}
\esube 
On the other hand, with the dissipaton decomposition
and its diffusion ansatz, \Eqs{patonF}--(\ref{gendiff_2}), together with the generalized Wick's theorem,\cite{Yan14054105,Wan22170901,Xu18114103}
we have
\bsube\label{appbcheck2}
\begin{align}
{\rm Tr}[\hat F^{\probe}(t- \tau)&
\hat q_1 \rho_{\T}(\tau)]
\nl =\sum_{kj}e^{-\gamma_{\tL k}(t-\tau)}&{\rm Tr}[\hat f_{0k}\hat f_{1j}\rho_{\T}(\tau)]
\nl = \sum_{kj}e^{-\gamma_{\tL k}(t-\tau)}& \sigma_{\tL k,\tR j}(\tau),
\end{align}
\begin{align}
{\rm Tr}[\hat F^{\probe}(t- \tau)&
\hat F \rho_{\T}(\tau)]=\Big({\rm Tr}[\hat F\hat F^{\probe}(t- \tau) \rho_{\T}(\tau)]\Big)^{\ast}
\nl = \sum_{kj}e^{-\gamma_{\tL k}(t-\tau)}&{\rm Tr}[\hat f_{0k}\hat f_{0j}\rho_{\T}(\tau)]
\nl = \sum_{kj}e^{-\gamma_{\tL k}(t-\tau)}& \big[\sigma_{\tL k,\tL j}(\tau)+\eta_{0k}\delta_{kj}\big]\nl = \sum_{kj}e^{-\gamma_{\tL k}(t-\tau)}& \sigma_{\tL k,\tL j}(\tau)+c_0(t-\tau),
\end{align}
\begin{align}
{\rm Tr}[\hat q_1^{\chain}(t- \tau)&
\hat q_1\rho_{\T}(\tau)]
=\Big({\rm Tr}[\hat q_1\hat q_1^{\chain}(t- \tau)\rho_{\T}(\tau)]\Big)^{\ast}
\nl = \sum_{kj}e^{-\gamma_{\tR k}(t-\tau)}&{\rm Tr}[\hat f_{1k}\hat f_{1j}\rho_{\T}(\tau)]
\nl = \sum_{kj}e^{-\gamma_{\tR k}(t-\tau)}& \big[\sigma_{\tR k,\tR j}(\tau)+\eta_{1k}\delta_{kj}\big]\nl = \sum_{kj}e^{-\gamma_{\tR k}(t-\tau)}& \sigma_{\tR k,\tR j}(\tau)+c_1(t-\tau),
\end{align}
and
\begin{align}
{\rm Tr}[\hat q_1^{\chain}(t- \tau)&
\hat F \rho_{\T}(\tau)]
\nl =\sum_{kj}e^{-\gamma_{\tR k}(t-\tau)}&{\rm Tr}[\hat f_{1k}\hat f_{0j}\rho_{\T}(\tau)]
\nl = \sum_{kj}e^{-\gamma_{\tR k}(t-\tau)}& \sigma_{\tL j,\tR k}(\tau).
\end{align}
\esube

Now let us check the consistence between \Eqs{appbcheck1}--(\ref{appbcheck2})
and \Eqs{b6a}--(\ref{b6c}).
Recall the initial state, $\rho_{\T}(t=0)\propto e^{-\beta_1 H_{\chain}}
\otimes
e^{-\beta_0 H_{\probe}}$, 
which leads to
$
    {\rm Tr}[\hat f_{r k} \hat f_{r' k'}\rho_{\T}(0)]=    \sigma_{r k, r' k'}(0)+\la\hat f_{r k} \hat f_{r' k'}\ra
$
thus $\sigma_{r k, r' k'}(0)=0$.
The solutions of \Eqs{b6a}--(\ref{b6c}) read
\bsube
\begin{align} 
   \sigma_{\tL k,\tL k'}(t) &= -i\alpha_1 \int_{0}^{t}\!{\rm d}\tau\,e^{-(\gamma_{\tL k} + \gamma_{\tL k'})(t-\tau)} 
   \nl & \quad \times 
   \sum_{j}\Big[(\eta_{0k}-\eta_{0\bar k}^{\ast}) \sigma_{\tL k',\tR j}(\tau)
   \nl & \quad\ \ 
   +(\eta_{0k'}-\eta_{0\bar k'}^{\ast}) \sigma_{\tL k,\tR j}(\tau)\Big], 
  \\ 
\sigma_{\tL k,\tR k'}(t) &=
-i\alpha_1 \int_{0}^{t}\!{\rm d}\tau\,e^{-(\gamma_{\tL k} + \gamma_{\tR k'})(t-\tau)}
\nl & \quad
\times \Big[(\eta_{0k}-\eta_{0\bar k}^{\ast})\sum_{j}\sigma_{\tR k',\tR j}(\tau)
\nl & \quad\ \ 
+ (\eta_{1k'}-\eta_{1\bar k'}^{\ast})\sum_{j}\sigma_{\tL k,\tL j}(\tau)
\nl & \quad\ \ 
+(\eta_{\tL k}\eta_{\tR k'} - \eta_{\tL \bar k}^*\eta_{\tR \bar k'}^*)\Big],
      \\
      \sigma_{1 k,1 k'}(t) &= -i\alpha_1  \int_{0}^{t}\!{\rm d}\tau\,e^{-(\gamma_{1 k} + \gamma_{1 k'})(t-\tau)} 
   \nl & \quad \times 
   \sum_{j}\Big[(\eta_{1k}-\eta_{1\bar k}^{\ast})\sigma_{0 j,1 k'}(\tau)
   \nl & \quad\ \ 
   +(\eta_{1k'}-\eta_{1\bar k'}^{\ast})\sigma_{0j,1 k}(\tau)\Big].
\end{align}
\esube
As a result,
\bsube
\begin{align} 
 &{\rm Tr}[\hat F^2\rho_{\T}(t)]= \sum_k \frac{1}{2}(\eta_{0k}+\eta^{\ast}_{0\bar k}) + \sum_{kk'}\sigma_{\tL k,\tL k'}(t) 
  \nl &
  = c_{0}(0)-2\alpha_1 \int_{0}^{t}\!{\rm d}\tau\,\phi_{0}(t-\tau)\sum_{kj}e^{-\gamma_{\tL k}(t-\tau)} 
   \sigma_{\tL k,\tR j}(\tau), \label{appFF}
   \\
    & {\rm Tr}[\hat F\hat q_1\rho_{\T}(t)]=\sum_{kk'}\sigma_{\tL k,\tR k'}(t) 
\nl & 
=-\alpha_1 \int_{0}^{t}\!{\rm d}\tau\,\phi_{0}(t-\tau)\sum_{kj}e^{-\gamma_{\tR k}(t-\tau)}
\sigma_{\tR k,\tR j}(\tau)
\nl &  \quad
-\alpha_1 \int_{0}^{t}\!{\rm d}\tau\,\phi_{1}(t-\tau)\sum_{kj}e^{-\gamma_{0 k}(t-\tau)}
\sigma_{0 k,0 j}(\tau)
\nl & \quad 
-i\alpha_1 \int_{0}^{t}\!{\rm d}\tau\,\big[c_0(t-\tau)c_1(t-\tau)-c^{\ast}_0(t-\tau)c^{\ast}_1(t-\tau)\big],
  \\ 
&{\rm Tr}[\hat q_1^2\rho_{\T}(t)]= \sum_k \frac{1}{2}(\eta_{1k}+\eta^{\ast}_{1\bar k}) + \sum_{kk'}\sigma_{\tR k,\tR k'}(t)  
  \nl & 
  =c_1(0) -2\alpha_1 \int_{0}^{t}\!{\rm d}\tau\,\phi_{1}(t-\tau)\sum_{kj}e^{-\gamma_{1 k}(t-\tau)} 
   \sigma_{0j,1k}(\tau),
\end{align}
\esube
which are easily found to be equivalent with \Eqs{appbcheck1}--(\ref{appbcheck2}).

\bibliographystyle{aiptit}
\bibliography{bibrefs}
\end{document}